\documentclass[runningheads]{llncs}
\usepackage{graphicx}
\usepackage{hyperref}
\begin{document}
\title{Urine Microscopic Image Dataset}
\author{Dipam Goswami\inst{1} \and
Hari Om Aggrawal\inst{2} \and
Rajiv Gupta\inst{3} \and
Vinti Agarwal\inst{1}}
% %
\authorrunning{D. Goswami et al.}
\institute{Birla Institute of Technology and Science Pilani, India \and
Independent Researcher, India. Previously with Institute of Mathematics and Image Computing, University of Lübeck, Germany \and Gurudwara Singh Sabha Charitable Dispensary, Rajpura, India}
% \renewcommand\UrlFont{\color{blue}\rmfamily}

\maketitle              % typeset the header of the contribution
%

% Birla Institute of Technology and Science Pilani, India,\\
% Independent Researcher, India. Previously with Institute of Mathematics and Image Computing, University of Lübeck, Germany.\\
% Gurudwara Singh Sabha Charitable Dispensary, Rajpura, India.

\begin{abstract}
Urinalysis is a standard diagnostic test to detect urinary system related problems. The automation of urinalysis will reduce the overall diagnostic time. Recent studies used urine microscopic datasets for designing deep learning based algorithms to classify and detect urine cells. But these datasets are not publicly available for further research. To alleviate the need for urine datsets, we prepare our urine sediment microscopic image (UMID) dataset comprising of around 3700 cell annotations and 3 categories of cells namely RBC, pus and epithelial cells. We discuss the several challenges involved in preparing the dataset and the annotations. We make the dataset publicly available\footnote{\url{https://github.com/dipamgoswami/UMID-Urine-Microscopic-Image-Dataset}}.

\end{abstract}
\section{Introduction}
Analysis of urine sediment particles from microscopic images play a vital role in the diagnosis of urinary and kidney diseases. The constituents of urine samples are Red Blood Cells (RBCs), White Blood Cells (WBCs) referred to as pus cells, epithelial cells, casts, bacteria, crystals and other artifacts. A routine urine test involves chemical, physical and microscopic analysis of the sample. We are concerned with the microscopic analysis of the urine sediment. Routine microscopic analysis generally looks for the presence of RBCs, pus cells and epithelial cells. A urine sample may also contain some indeterminate objects as well as some out-of-focus objects and lens artifacts. 

The classification of cells is vital for the detection of diseases and its diagnosis. This work is aimed at providing medical support to medical centres across rural areas where there is a shortage of skilled lab technicians to analyze urine samples of patients. To the best of our knowledge, our UMID dataset is the first urine images dataset to be made publicly available to all researchers. Our dataset has been annotated under the supervision of medical professionals and have been cross-verified. We expect that the availability of the dataset will accelerate research in the urine sediment cell detection domain.

\section{Related Work}
\textbf{Urine Sediment Datasets} Various datasets have been used to study cell detection from urine sediments though not publicly available. Some of these studies like~\cite{Xu2019} are focused on classifying only RBC and WBC cells while other works~\cite{Liang2018} classified objects into erythrocyte, leukocyte, epithelial cell, crystal, cast, mycete, epithelial nuclei and noise,~\cite{Li2020} used sub-classes of epithelial cells like low-transitional epithelium and squamous epithelial cells and~\cite{Ji2019} also considered bacteria and sperms. Some papers~\cite{Ji2019,Aziz2018} have highlighted the regular occurrence of WBC and RBC clusters in urine datasets but are limited to detecting the clusters~\cite{Ji2019} or predicting segmentation mask boundaries for RBC and WBC clumps~\cite{Aziz2018}. 

Detecting individual cells from these cell clusters have not been studied yet and our work focuses to detect the densely packed cells using weak instance-level information. With improved precision in detection of cluster cells, the urine test reports can give a better count of the different cells leading to better diagnosis of diseases.
In our dataset, we have observed RBCs, pus cells and epithelial cells as the main components. We have labelled the confusing and poorly focused components into a missedlabel category and we avoid these instances while training the deep learning models.

\section{UMID Dataset}
A physician at our university hospital collected all urine microscopic images in the course of his patient diagnosis. He followed the normal routine procedure for data collection. His primary objective was to differentiate and count urine sediments manually from the images for diagnosis instead of acquiring good quality images for the dataset. Hence, the acquired images is not a best set of images that one would like to have to train a machine learning model in view of obtaining a remarkable performance. But, the acquired dataset aligns quite well with our objective. In this study, our goal is to build a neural network model that generalizes well and work satisfactorily even with low quality images. We explain image quality in our context later in this section. 

In practice, doctors and lab technicians does not follow the recommended protocols \cite{Bunjevac2017} very precisely due to the shortage of time specially in highly populated countries where the daily influx of patients is quite high. Moreover, microscopes does not get servicing regularly that leads to many flaws in the images.
Instead of assuming ideal situation, we should design a robust learning framework that could sustain these issues up to some extent. 
All the automatic urine analyzers comes with a special microscope that is attached to the analyzers. Hence, they works well in practice.
However, our long term goal is to develop a system that works with existing lab setup with only a minor addition of camera or a smartphone to one of the eyepiece.

Due to the unavailability of proper infrastructure, the patient information is not stored anywhere. Hence, images are inherently anonymized at the source. Images are subject to bias of the doctor towards his acquisition habits; it is one of the limitation of our study.

\subsection{Urine sediments and their clinical significance}

A routine urine diagnostic report mention the numbers of erythrocytes/RBC, leukocytes/pus,
epithelial cells, and cast per high power field of a urine sample. Other urine sediment constituents such as crystals, bacteria, yeast cells, salts, and spermatozoa are not counted, but instead given as crosses \cite{Neuendorf2020}. Casts appears very rarely in a urine sample \cite{Neuendorf2020}. During our data collection duration, we did not get any urine sample with cast. Hence, in this study our focus is mainly to detect and classify only three types of urine sediments, RBC, pus, and epithelial cell. These three classes can be also sub classified. For example, the epithelial cells can further be classified into squamous and transitional epithelial cells. Though, these sub-classifications have clinical relevance, but they are hardly reported in a routine urine diagnostic report. 

\subsubsection{Morphology}
RBC are round cells, contains no nucleus or granules, smaller than pus cells, and has relatively sharp boundaries. Based on the pH value of the urine, it is either biconcave, disc-shaped (pH = 6), thorn apple-shaped (pH $<$ 6), or a light disc (pH $>$ 6). They also sometimes appears as ring-shaped. Pus cells are generally larger than RBC cells. They are round in shape and has dark and granular surface. Pus cells could also have segmented nucleus. Pus cells lie individually and in clusters. Epithelial cells are irregular in shape, generally larger than pus cells, and has a small size nucleus. Sometimes, epithelial cells have two nucleus as well. These cells could also lie individually and clusters.

\subsubsection{Clinical significance}
The normal range for RBCs are 0-1/HPF. Higher values of RBC indicates hematuria in kideny tumors, urinary tract tumors, urinary tract infections, or trauma. The normal range of pus cells are 1-4/HPF. Excess of it could indicates inflammation in the urinary tract system and inflammatory renal disease.
The normal range for epithelial cells are 0-15/HPF. A higher value indicates inflammation of urinary tract system.

\subsection{Image acquisition - \\ recommended vs in practice} \label{subsec:image_acquisition}
We used two bright-field microscopes for acquiring images. Images from one microscope was of a lower quality than the other. Majorly, we observe a half circular ring artifact in the images. We could not find the exact reason behind the artifact, but we suspect that it is mainly because of misalignment between lens and the light source. In our dataset, the ratio of images are approximately 10:2 from old to new microscopes. We could acquire only a limited amount of images from the hospital due to the limited patients visiting the hospital due to the COVID-19 restrictions. 

The urine sample on a glass slide has a certain depth and the cells are present at different depths. Hence, we say that the urine slide has a multi-layer structure. Due to that, all the cells are not clearly visible within the depth of field of a lens focused at a particular focal plane. Hence, numerous focus adjustments are required to identify every cell in the sample. Moreover, microscope could examine only a certain area of the whole urine sample. Hence, it is recommended to examine 20-30 high power fields (HPF) for one urine sample and accumulate the findings to build the final diagnostic report. Certainly, it is a very time consuming task. 

We observed in practice that the doctors follow the recommendations only for a few sensitive cases. Otherwise, doctors spend only a few seconds to analyse one urine slide that covers approximately 10 HPF in a urine slide and adjust focus of the lens up to a certain level that is enough to differentiate the cell features and its size. Hence, the acquired images majorly have a certain amount of blur and hence, object edges are not very sharp and cell granularity deteriorates. 

There a few more limitations that originates from urine multi-layer structure. We observe that cells in an image can have a different degree of blur based on their depth with respect to the distance from the lens. Few cells could be highly out of focus that lies beyond the depth-of-field of the lens. This irregular blurring effect makes it very hard sometimes to differentiate cells even for an experienced doctor. To handle such cases, we introduced a missed label class.

Both microscopes are the product of MAGNUS OPTO SYSTEMS INDIA PVT. LTD, model number CH20iBIMF. Eyepieces are with 10x magnification and 18mm field number and the achromatic objective lens are with 40x magnification. Each high power field is examined at 400×magnification (corresponding to an 10× eyepiece and a 40× objective). We connected a COSLAB 5 megapixel microscope digital camera (model number COSUSB5000) to one of the eyepiece of the microscope to acquire a digital image of the urine sample and store it on the desktop. The images are acquired at the lower resolution of $1280 \times 720$. The video feed to the desktop from the camera was quite slow at higher resolution and hence avoided by the doctor for image acquisition. If camera is attached to the microscope, doctors prefer to use video feed on the desktop to differentiate cells rather than directly observing through the eyepiece. The video feed provides more flexibility and easiness to scan the entire urine slides.

%Microscope digital camera: COSLAB COSUSB5000 5.0MP 1/2.5" APTINA CMOS
%Eyepiece 10X/18 
%Lens Magnus i NEA 40X/0.65, 160/0.17
%Microscope: Model CH20iBIMF, MAGNUS OPTO SYSTEMS INDIA PVT. LTD.
%Camera:

%A semi-quantitative evaluation is carried out by examining 20–30 high power fields (HPF) in a meandering pattern at 400×magnification (corresponding to an 10× eyepiece and a 40× objective) while considering a certain field number.

%Irrespective of the magnification used, micrometer fine-focus adjustment must be constantly operated in order to fine-tune the microscopic plane to ensure that no components are overlooked.

% image resolution and interpolation

\subsection{Novel data annotation strategies}
% blur, low contrast
% multi-plane
% overlapping of cells, no regular pattern observed

Annotating a urine microscopic image is a challenging task. Optical blurring, overlapping of cells, low contrast, small cell size are some sources of difficulties. It is easy to overlook RBC cells due to their small size and very smooth surface that leads to very low contrast in the surrounding of the cell.

Three authors of this paper including a doctor jointly annotated the cells in the images. We used Microsoft VoTT software for annotations. We primarily followed three strategies to annotate cells from which cluster and missed label strategies are novel approaches that we proposed in this work. These novel approaches are based on the properties of urine microscopic images that we describe in the next paragraph. Moreover, these approaches greatly reduces the overall annotation time. Doctors are generally over occupied with patients and have very busy schedule. Hence, reducing the annotation time greatly helped to get feedback from the doctors.

%with proper training one can perform annotation..
%doctors are confused ...
%It is difficult to get hold on doctors for annotating images due to their very busy schedules. It was the primary motivation for us to reduce annotation time by exploiting the similarly between each cell class. 

\subsubsection{Standard annotation strategy}
For a object detection and classification task, the standard approach is to train the neural network with respect to the ground truth of all the foreground objects present in the image. Each object is assigned a class and a bounding box that encodes the location and size of the object in a given image. We follow the same annotation strategy in our work, but only for a few cells that lie individually in the image.

\subsubsection{Novel missed label class}
% include optical blurring
In some circumstances, doctors could not make a reliable guess about the appropriate class of a cell. Hence, in theory, we can not place them neither in main classes nor in the background. If we treat them as a background, in principle, we are forcing network to learn a wrong classification. Instead of that, we place such cells in a separate missed label class and do not include members of this class during training and leave it to the network to predict the class for such cells.

\subsubsection{Point annotations for clusters}\label{sec:ptsForClusters}
In urine images, cells are also found in clusters. In clusters, cells are densely packed and overlapping with each other, hence, annotation is a difficult, puzzling, and a time-consuming task inside a cluster. The annotations can result in inconsistent bounding boxes. Hence, we avoid annotating each cell inside the clusters.

We propose to specify a single point on the cell (near to its center approximately) instead of deciding boundaries and drawing boxes to localize the cells in a cluster. The size of cells can be estimated from points and these generated pseudo boxes can be used during training. Annotating a point is a much simpler task and reduces time, efforts, and dilemma.

\subsection{Data statistics and analysis}

\subsubsection{Training, testing, and validation set}

\begin{figure}[htp]
    \centering
    \includegraphics[width=12cm]{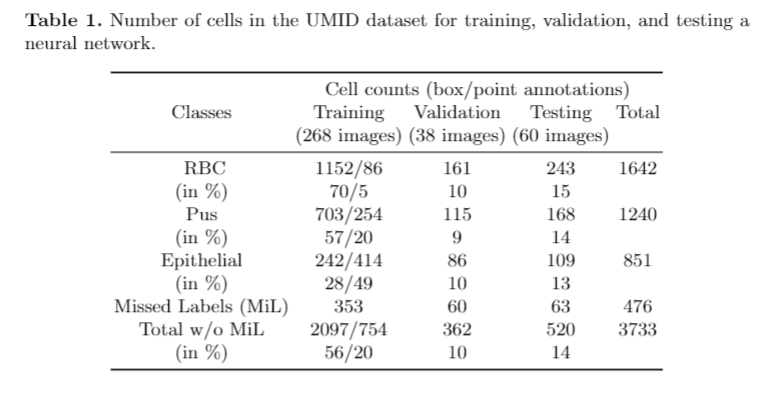}
    \label{fig:data}
\end{figure}

In the UMID dataset, we have 366 urine images of resolution $1280 \times 720$ that we divided into three sets for training ($\approx 76\%$), validation ($\approx 10\%$), and testing ($\approx 14\%$); see Table~\ref{tab:cell_stats}. The division is based on the total 3733 annotated cells in the dataset instead of the images. The cell features are the primary source of learning.

As discussed in the last section, cells inside the cluster are difficult to annotate; hence we annotate cells with boxes (outside clusters) and points (inside clusters) in the training dataset. But for testing and validation dataset, we purposely annotate cells lies inside clusters with boxes to evaluate the detection performance of the ML model. Generally, pus and epithelial cells form clusters; hence, we have more points annotations compared to the RBC cells. In total, $20\%$ cells are annotated with points for which pseudo ground truth boxes are required for training.

The dataset has approximately $44\%$ RBC, $33\%$ pus, and $23\%$ epithelial cells. Though, the classes are slightly imbalanced, the performance of the ML model is not biased towards one class.

We have approximately $11\%$ cells for which the annotators were unsure about the correct class. For these cells only bounding boxes are drawn and label them to missed label class. These structures are not used during training the models.

\bibliographystyle{IEEEtran}
\bibliography{main}

\begin{thebibliography}{1}
\providecommand{\url}[1]{#1}
\csname url@samestyle\endcsname
\providecommand{\newblock}{\relax}
\providecommand{\bibinfo}[2]{#2}
\providecommand{\BIBentrySTDinterwordspacing}{\spaceskip=0pt\relax}
\providecommand{\BIBentryALTinterwordstretchfactor}{4}
\providecommand{\BIBentryALTinterwordspacing}{\spaceskip=\fontdimen2\font plus
\BIBentryALTinterwordstretchfactor\fontdimen3\font minus
  \fontdimen4\font\relax}
\providecommand{\BIBforeignlanguage}[2]{{%
\expandafter\ifx\csname l@#1\endcsname\relax
\typeout{** WARNING: IEEEtran.bst: No hyphenation pattern has been}%
\typeout{** loaded for the language `#1'. Using the pattern for}%
\typeout{** the default language instead.}%
\else
\language=\csname l@#1\endcsname
\fi
#2}}
\providecommand{\BIBdecl}{\relax}
\BIBdecl

\bibitem{Xu2019}
\BIBentryALTinterwordspacing
X.-T. Xu, J.~Zhang, P.~Chen, B.~Wang, and Y.~Xia, ``Urine sediment detection
  based on deep learning,'' in \emph{Intelligent Computing Theories and
  Application}.\hskip 1em plus 0.5em minus 0.4em\relax Springer International
  Publishing, 2019, pp. 543--552. [Online]. Available:
  \url{https://doi.org/10.1007/978-3-030-26763-6_52}
\BIBentrySTDinterwordspacing

\bibitem{Liang2018}
\BIBentryALTinterwordspacing
Y.~Liang, R.~Kang, C.~Lian, and Y.~Mao, ``An end-to-end system for automatic
  urinary particle recognition with convolutional neural network,''
  \emph{Journal of Medical Systems}, vol.~42, no.~9, Jul. 2018. [Online].
  Available: \url{https://doi.org/10.1007/s10916-018-1014-6}
\BIBentrySTDinterwordspacing

\bibitem{Li2020}
\BIBentryALTinterwordspacing
Q.~Li, Z.~Yu, T.~Qi, L.~Zheng, S.~Qi, Z.~He, S.~Li, and H.~Guan, ``Inspection
  of visible components in urine based on deep learning,'' \emph{Medical
  Physics}, vol.~47, no.~7, pp. 2937--2949, May 2020. [Online]. Available:
  \url{https://doi.org/10.1002/mp.14118}
\BIBentrySTDinterwordspacing

\bibitem{Ji2019}
\BIBentryALTinterwordspacing
Q.~Ji, X.~Li, Z.~Qu, and C.~Dai, ``Research on urine sediment images
  recognition based on deep learning,'' \emph{{IEEE} Access}, vol.~7, pp.
  166\,711--166\,720, 2019. [Online]. Available:
  \url{https://doi.org/10.1109/access.2019.2953775}
\BIBentrySTDinterwordspacing

\bibitem{Aziz2018}
\BIBentryALTinterwordspacing
A.~Aziz, H.~Pande, B.~Cheluvaraju, and T.~R. Dastidar, ``Improved extraction of
  objects from urine microscopy images with unsupervised thresholding and
  supervised u-net techniques,'' in \emph{2018 {IEEE}/{CVF} Conference on
  Computer Vision and Pattern Recognition Workshops ({CVPRW})}.\hskip 1em plus
  0.5em minus 0.4em\relax {IEEE}, Jun. 2018. [Online]. Available:
  \url{https://doi.org/10.1109/cvprw.2018.00299}
\BIBentrySTDinterwordspacing

\bibitem{Bunjevac2017}
\BIBentryALTinterwordspacing
A.~Bunjevac, N.~N. Gabaj, M.~Miler, and A.~Horvat, ``Preanalytics of urine
  sediment examination: effect of relative centrifugal force, tube type, volume
  of sample and supernatant removal,'' \emph{Biochemia Medica}, vol.~28, no.~1,
  Dec. 2017. [Online]. Available: \url{https://doi.org/10.11613/bm.2018.010707}
\BIBentrySTDinterwordspacing

\bibitem{Neuendorf2020}
\BIBentryALTinterwordspacing
J.~Neuendorf, \emph{Urine Sediment}.\hskip 1em plus 0.5em minus 0.4em\relax
  Springer International Publishing, 2020. [Online]. Available:
  \url{https://doi.org/10.1007/978-3-030-15911-5}
\BIBentrySTDinterwordspacing

\end{thebibliography}
\end{document}